\documentclass[authoryear,preprint,review,12pt]{elsarticle}

\usepackage{amsmath}
\usepackage{amssymb}
\usepackage{graphicx}
\usepackage{natbib}
\usepackage{color}
\usepackage{subfigure}

\journal{Journal of Theoretical Biology}

\usepackage{url}

\newcommand{\x}{\mathbf{X}}
\newcommand{\V}{n}
\newcommand{\lib}{L}
\newcommand{\U}{\mathbf{U}}
\newcommand{\E}{\mathbf{E}}
\newcommand{\var}{\sigma}
\newcommand{\p}{u}
\newcommand{\qi}{v}
\newcommand{\qij}{w}
\newcommand{\cp}{q}

\newcommand{\multinom}[2]{\genfrac{[}{]}{0pt}{}{#1}{#2}}

\DeclareMathOperator{\Cov}{Cov}

\begin{document}

\begin{frontmatter}

\title{Calculating complexity of large randomized libraries}
\author{Yong Kong}
\address{
Department of Molecular
Biophysics and Biochemistry\\
W.M. Keck Foundation Biotechnology Resource Laboratory \\
Yale University\\
333 Cedar Street, New Haven, CT 06510 \\
\ead{yong.kong@yale.edu}
}

\begin{abstract}
Randomized libraries are increasingly popular in protein engineering
and other biomedical research fields.
Statistics of the libraries are useful 
to guide and evaluate randomized library construction.
Previous works only give the mean of the number of unique sequences
in the library, and they can only handle equal molar ratio of 
the four nucleotides at a small number of mutation sites.
We derive formulas to calculate the mean and variance
of the number of unique sequences in 
libraries generated by cassette mutagenesis
with  mixtures of arbitrary nucleotide ratios.  
Computer program was developed which utilizes arbitrary numerical precision
software package 
to calculate the statistics of large libraries.
The statistics of library with mutations 
in more than $20$ amino acids can be calculated easily. 
Results show that the nucleotide ratios have significant effects
on these statistics.  
The more skewed the ratio, the larger the library size is needed 
to obtain the same expected number of unique sequences.
The program is freely available at 
\url{http://graphics.med.yale.edu/cgi-bin/lib_comp.pl}.

\end{abstract}

\end{frontmatter}

\section{Introduction}

Randomized libraries are widely used to 
select novel proteins with specific biological and physiochemical
properties.
There are many protocols to create randomized libraries
\citep{Neylon2004},
such as cassette mutagenesis and error-prone PCR (epPCR).
In the cassette mutagenesis,
random mutagenesis is generated in
a particular region or regions of the target gene
through incorporation of degenerate synthetic DNA sequence.
Usually equal molar nucleotides are used ($1:1:1:1$ ratio of nucleotides),
but on other occasions,
different molar ratios of nucleotides in the mixtures are used
to create predetermined physiochemical properties
in the targeted region(s).
For example, 
in the study of selection and characterization of small transmembrane
proteins that bind and activate the 
platelet-derived growth factor $\beta$ (PDGF $\beta$) receptor,
fifteen transmembrane amino acids of E5 protein of bovine papillomavirus (BPV)
were replaced with random sequences~\citep{Freeman-cook2004}.
The following library design was used to mimic the hydrophobic transmembrane
region of BPV E5 protein:
\begin{equation} \label{E:s1}
  \cdots \text{(NXR)}_3 \text{CAA} \text{(NXR)}_{12} \cdots 
\end{equation}
where the three \verb+NXR+ codons are followed by a \verb+CAA+ codon encoding
glutamine, which in turn is followed by 12 more \verb+NXR+ codons.
For the \verb+NXR+ codons,
\verb+N+ stands for an equal mix of \verb+A+, \verb+C+, \verb+G+, and \verb+T+,
\verb+X+ is a $5:0.1$ mixture of  \verb+T:C+,
and \verb+R+ is an equal mix of \verb+A+ and \verb+G+.

In order to guide and evaluate randomized library construction,
the statistical properties of the libraries will be useful.
One of the most asked questions is the number of unique sequences
in the library (the complexity of the library). 
In the following we are going to give some formulas to calculate
the expected number of unique sequences in the library
as well as its variance.
Our treatment is different from previous works in several ways 
\citep{Bosley2005,Firth2008}.
Firstly, the previous works deal with only mixtures of
equal molar ratio of the four nucleotides, 
while we can handle mixtures of arbitrary user-defined molar ratios,
which is more useful in such situations as described above.
As shown in the examples below, the different molar ratios
in the nucleotide mixtures make a significant difference
in the statistics of the library.
Secondly, we present a formula for the variance, and hence the standard
deviation, of the number of unique sequences in the library.
The standard deviation gives an indication of the spread of the
distribution around the expected value.
Thirdly, by using a mathematical software library that can handle
arbitrary numerical precision, we can calculate the statistics
for much larger libraries. The statistics of library with mutations 
in more than $20$ amino acids can be calculated easily. 
The program can be accessed freely in
the web server at 
\url{http://graphics.med.yale.edu/cgi-bin/lib_comp.pl}.

The paper is organized as follows.
In the \emph{Theory} section we derive the formulas for the
expected number of unique sequences
in the randomized library and its variance.
Within the assumption of the model, the formulas are exact.
For real calculations of randomized libraries, however,
these formulas have to be rearranged due to the huge number of
possible sequences.
In \emph{Software implementation} section we discuss
several ways to make the calculation manageable while keeping the
numerical accuracy of the calculation.
In the last section two examples are given for a small library
and a relatively bigger library.

\section{Theory}

Assume that the size of the library (the number of transformants) 
is $\lib$ and
the total number of all possible sequences is $\V$.
Usually $\V$ is a huge number for a large randomized library.
For example, if $60$ nucleotide bases are mutated,
the number of potential sequences is
$\V = 4^{60} \approx 1.3 \times 10^{36}$.
We denote the probability of sequence $s_i$ as $p_i$.
For each sequence $s_i$ among these possible sequences,
we can associate a random variable $\x_i$, which is either $1$ or $0$,
according as the sequence $s_i$ is or is not in the library.
The respective probabilities of $\x_i$ to take these two values are:
\begin{subequations} \label{E:xi}
\begin{align}
  \Pr \{ \x_i = 0 \} &= (1-p_i)^\lib      = \qi_i ,\\
  \Pr \{ \x_i = 1 \} &= 1 - (1-p_i)^\lib  = \p_i .
\end{align} 
\end{subequations}
The \emph{number of unique sequences} in the library is given by the
random variable $\U_n$:
\[
 \U_n = \x_1 + \x_2 + \cdots + \x_n = \sum_{i=1}^n \x_i.
\]

The properties that we are interested in are the average (expectation) of
$\U_n$ and its variance.
The expectation gives 
the average of the number of distinct sequences in the library, 
and the variance shows the tightness of the distribution of
the number of distinct sequences around its average.
The average is given by
the expectation of $\U_n$ as
\begin{equation}
 \E(\U_n) = \sum_{i=1}^n \E(\x_i),
\end{equation}
and its variance is given by
\begin{equation}
 \var^2(\U_n) = \sum_{i=1}^n \var^2(\x_i) + 2 \sum_{i>j}\Cov(\x_i, \x_j),
\end{equation}
where $\Cov(\x_i, \x_j)$ is the \emph{covariance} of $\x_i$ and $\x_j$:
\begin{align*}
 \Cov(\x_i, \x_j) &= \E( (\x_i - \E(\x_i) ) (\x_j - \E(\x_j) ) \\
                  &= \E( \x_i \x_j) - \E( \x_i) \E(\x_j)  .
\end{align*}

From Eq.~\eqref{E:xi} we know that both $\E(\x_i)$ and $\E(\x_i^2)$ 
equal to $\p_i$:
$
 \E(\x_i) = 0 \cdot \qi_i + 1 \cdot \p_i = \p_i,
$
$
 \E(\x_i^2) = 0^2 \cdot \qi_i + 1^2 \cdot \p_i = \p_i .
$
Hence the variance of $\x_i$ is
\[
 \var^2(\x_i) = \E(\x_i^2) - \E(\x_i)^2 = \p_i \qi_i  .
\]
To calculate $\var^2(\U_n)$ we need $\Cov(\x_i, \x_j)$, which in turn 
depends on $\E( \x_i \x_j)$.
The joint probability distribution of $\x_i$ and $\x_j$ are given by
\begin{align*}
  \Pr \{ \x_i = 0, \x_j = 0 \} &= (1-p_i-p_j)^\lib = \qij_{ij}         ,  \\
  \Pr \{ \x_i = 0, \x_j = 1 \} &= (1-p_i)^\lib - (1-p_i-p_j)^\lib 
                                = \qi_i - \qij_{ij},  \\
  \Pr \{ \x_i = 1, \x_j = 0 \} &= (1-p_j)^\lib - (1-p_i-p_j)^\lib   
                                = \qi_j - \qij_{ij}    ,  \\
  \Pr \{ \x_i = 1, \x_j = 1 \} &= 1 - (1-p_i)^\lib - (1-p_j)^\lib \notag \\ 
  + (1-p_i-p_j)^\lib           &= 1 - \qi_i - \qi_j + \qij_{ij} .
\end{align*} 
From the joint probability the expectation of $\x_i \x_j$ can be obtained as
\begin{align*}
 \E( \x_i \x_j) &= 1 - (1-p_i)^\lib - (1-p_j)^\lib 
  + (1-p_i-p_j)^\lib \\
 &= 1 - \qi_i - \qi_j + \qij_{ij}
\end{align*}
from which we obtain the covariance of $\x_i$ and $\x_j$ as
\[
 \Cov(\x_i, \x_j) = \qij_{ij} - \qi_i \qi_j .
\]
Putting all the pieces together we have the average and variance of
$\U_n$ as
\begin{equation} \label{E:avg}
 \E(\U_n)= \sum_{i=1}^n \p_i = \sum_{i=1}^n \left[ 1 - (1-p_i)^\lib   \right], 
\end{equation}
and
\begin{align}    \label{E:var}
 \var^2(\U_n) &= \sum_{i=1}^n \p_i \qi_i + 2 \sum_{i>j} \left[ \qij_{ij} - \qi_i \qi_j \right]  \notag \\
 &= \sum_{i=1}^n \qi_i - \left[ \sum_{i=1}^n \qi_i \right]^2
 + 2 \sum_{i>j} \qij_{ij} \notag \\
 &= \sum_{i=1}^n (1 - p_i)^\lib 
 - \left[ \sum_{i=1}^n (1 - p_i)^\lib  \right]^2
 + 2 \sum_{i>j} (1 - p_i - p_j)^\lib   .
\end{align}

\section{Software implementation}

When the number of possible sequences $n$ is small, the average and variance
of the unique sequences in the library can be calculated directly using
Eqs.~\eqref{E:avg} and \eqref{E:var}.
When $n$ becomes large, however, a direct calculation is not feasible.
If in one position along the sequence we have $m$ possibilities (for nucleotide
$m$ usually is $4$), and we have mutation in $b$ such positions,
a direct calculation would have $n = m^b$ possible values of $p_i$,
which, as stated earlier, is too big to tackle directly.
Not all these $p_i$, however, are unique.
For a particular mixture ratio of \verb+X+,
we can calculate
the probability $\cp_i$ of each nucleotide in that position 
from the nucleotide ratio,
which is just the fraction of each nucleotide at 
positions with mixture ratio \verb+X+.
For example, if the ratio is $1:2:3:4$ for
\verb+A+, 
\verb+C+, 
\verb+G+, 
and \verb+T+,
then $\cp_i$ will take values of $1/10$, $2/10$, $3/10$, and $4/10$.
All possible $p_i$ are given in the following multinomial expansion
\begin{equation} \label{E:expmb}
 (\cp_1 + \cp_2 + \cdots + \cp_m)^b
 = \sum_{\sum i_j = b} \multinom{b}{i_1, i_2, \dots, i_m} 
 \cp_1^{i_1} \cp_2^{i_2} \cdots \cp_m^{i_m} .
\end{equation}
For each possible $p_i = \cp_1^{i_1} \cp_2^{i_2} \cdots \cp_m^{i_m}$, 
the number of such $p_i$ is given by the multinomial coefficient,
\[
 c_i = c(i_1, i_2, \dots, i_m) = \multinom{b}{i_1, i_2, \dots, i_m} 
 =  \frac{b!}{i_1! i_2! \cdots i_m!} .
\]
The number of such unique probabilities,
which equals the number of unique terms in the expansion of 
Eq.~\eqref{E:expmb},  is given by
\begin{equation} \label{E:h}
 h =  \binom{b + m - 1}{m-1} = \frac{(b+m-1)!}{b! (m-1)!},
\end{equation}
which is much smaller than $m^b$.

Hence the mean in Eq.~\eqref{E:avg} can be rewritten as
\begin{equation} \label{E:avg2}
 \E(\U_n)= \sum_{j=1}^h c_j \p_j
\end{equation}
and the variance in Eq.~\eqref{E:var} as
\begin{multline} \label{E:var2}
 \var^2(\U_n) = \sum_{i=1}^h c_i \qi_i 
 - \left[  \sum_{i=1}^h c_i \qi_i \right]^2 \\
 + 2 \left[  \sum_{i=1}^h \frac{c_i (c_i - 1)}{2} w_{ii} 
   + \sum_{i>j} c_i c_j w_{ij} \right] .
\end{multline}
Furthermore, the number of the terms in 
Eqs.~\eqref{E:avg2} and \eqref{E:var2}, $h$,
 can be reduced
if there is degeneracy among $\cp_i$:
\[
 (\cp_1 + \cp_2 + \cdots + \cp_m)^b = 
(\alpha_1 \cp_1' + \alpha_2 \cp_2' +  \cdots + \alpha_{m'} \cp_{m'}')^b.
\]
where we combine terms of $q_i$ of the same value together (there are
$\alpha_i$ of them):
 $\cp_i'$ are just unique items among $\cp_i$.
In such cases the number of unique probabilities of $p_i$ is given by
\begin{equation} \label{E:h'}
 h' =  \binom{b + m' - 1}{m'-1} = \frac{(b+m'-1)!}{b! (m'-1)!} .
\end{equation}
For example, if we want to mutate nucleotide acids in $100$ positions with
nucleotide ratio $1:2:1:2$, then $b=100$, $m=4$ and $m'=2$. 
The brute force calculation would have to add up $m^b = 1.6 \times 10^{60}$
terms.
Eq.~\eqref{E:h} gives $h=176851$, while Eq.~\eqref{E:h'} gives $h'=101$,
a significant reduction.

The above statements apply to mutations with a single base composition, as in
$X X \cdots X = (X)_b$.  They can, however, be generalized easily to handle
multiple base compositions,
as in $(X_1)_{b_1} (X_2)_{b_2} \cdots (X_k)_{b_k}$. 
For example,
the mutation in \eqref{E:s1} 
can be written as
\[
 N_{15} X_{15} R_{15}
\]
for the computation purpose.
In this general situation, the unique probabilities $p_i$ and its
associated coefficient $c_i$ are given by the expansion of
the product
\begin{equation} \label{E:expan}
  \prod_{i=1}^k (\cp_{i1} + \cp_{i2} + \cdots + \cp_{im_i})^{b_i} = 
  \prod_{i=1}^k (\alpha_{i1} \cp_{i1}' + \alpha_{i2} \cp_{i2}' 
  +  \cdots + \alpha_{im_i'} \cp_{im_i'}')^{b_i} .
\end{equation}

A C program has been written that uses Eqs~\eqref{E:avg2} and \eqref{E:var2}
to calculate the average and variance of the number of
unique sequences in randomized libraries,
where $c_i$ and $p_i$ are from the expansion of Eq.~\eqref{E:expan}. 
For small libraries, numerical accuracy is not a problem: 
standard programming languages are sufficient to give correct answers.
For large libraries, careful attention has to be paid to the numerical
stability, since there are many terms involved, and each term of probability
$p_i$ is very small, as $p_i$ is the product of many $\cp_j$ raised
to high power. Standard programming languages like C usually cannot
handle the situation well.

To overcome the issue of numerical accuracy, the program links the library
of PARI/GP package
\citep{PARI2008}, 
which can do numerical calculations with arbitrary
precisions.
Furthermore, the package has the ability to do symbolic calculations,
which makes some of the above mentioned calculations easier.
A web server has been set up to access the program 
at \url{http://graphics.med.yale.edu/cgi-bin/lib_comp.pl}. 

The program is simple to use.  
The user just type in the library size $\lib$ and 
the nucleotide ratios of the mixtures, 
followed by the number of bases with that nucleotide ratio, 
separated by a space.
Multiple ratios of the mixtures can be handled.
If one or more of the nucleotides is not included at a position,
simply exclude them in the input
(although including them as zeros does not hurt: the program filters
them out automatically).
For example,
to calculate the complexity for the library as described
in Eq.\eqref{E:s1} in the \emph{Introduction} section,
user input for the nucleotide ratios and base numbers is in the format of:
\begin{equation}
 \begin{verb}
   1:1:1:1 15   5:0.1 15   1:1 15  
 \end{verb}
\end{equation}

\section{Examples}
In this section two artificial examples are given to show the effects
of different molar ratios of the nucleotide mixtures on the statistics
of the library.
The first example is a small randomized library with mutations
on two amino acids, with different nucleotide mixture ratios
for the first, second, and third codon positions: $(X)_2(Y)_2(Z)_2$.
Here we have $k=3$ and $b_1 = b_2 = b_3 = 2$.
The potential number of sequences is $m^b = 4^6 = 4096$.
The second library is relatively larger, with mutations in
$8$ amino acids: $(X)_8(Y)_8(Z)_8$.
Here we have $k=3$ and $b_1 = b_2 = b_3 = 8$.
The potential number of sequences is $m^b = 4^{24} \approx 2.8 \times 10^{14} $.
Three different sets of molar ratios are used for each library:
\begin{align*}
 &\text{Set 1:} &\text{equal ratio} &     &\\
 &\text{Set 2:} & X=1:1:1:8,        &Y=1:1:1:9,  &Z=1:1:1:10\\
 &\text{Set 3:} & X=1:1:1:10,       &Y=1:1:1:30, &Z=1:1:1:40\\
\end{align*}
Set 3, though not very common in practice,
is included to show the effects of nucleotide molar ratio on the
library statistics.
 
\begin{figure}[!tpb]
  \centering
  \subfigure[The expectation of the number of unique sequences in the library.]
	    {
	     \includegraphics[angle=270,width=\columnwidth]{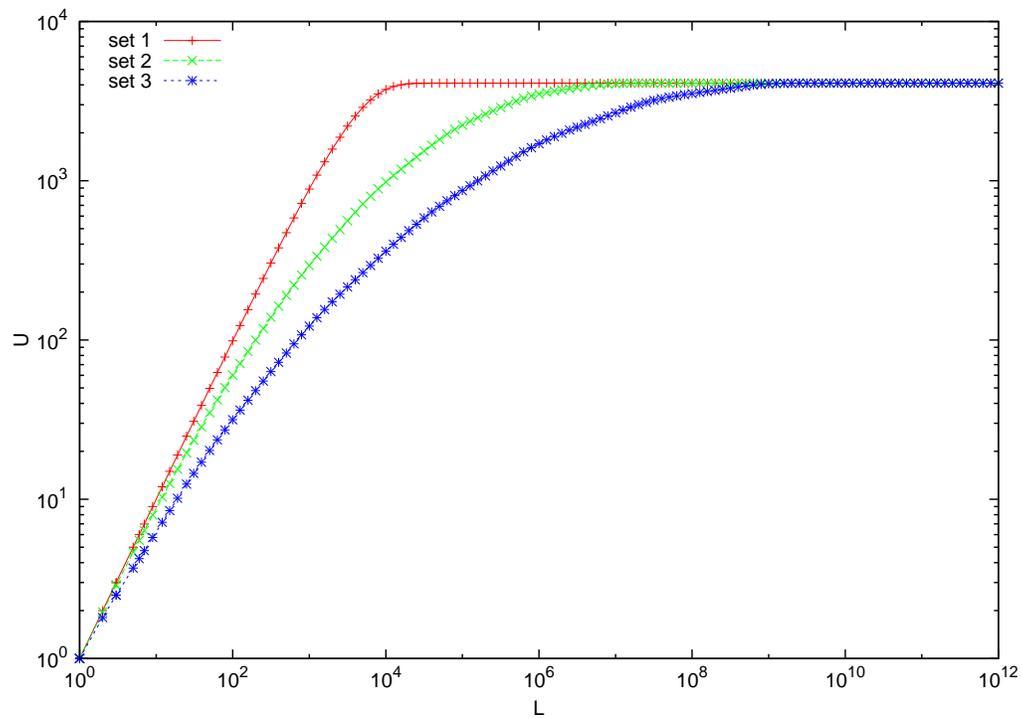}
	    }
  \subfigure[The standard deviation.]
	    {
	     \includegraphics[angle=270,width=\columnwidth]{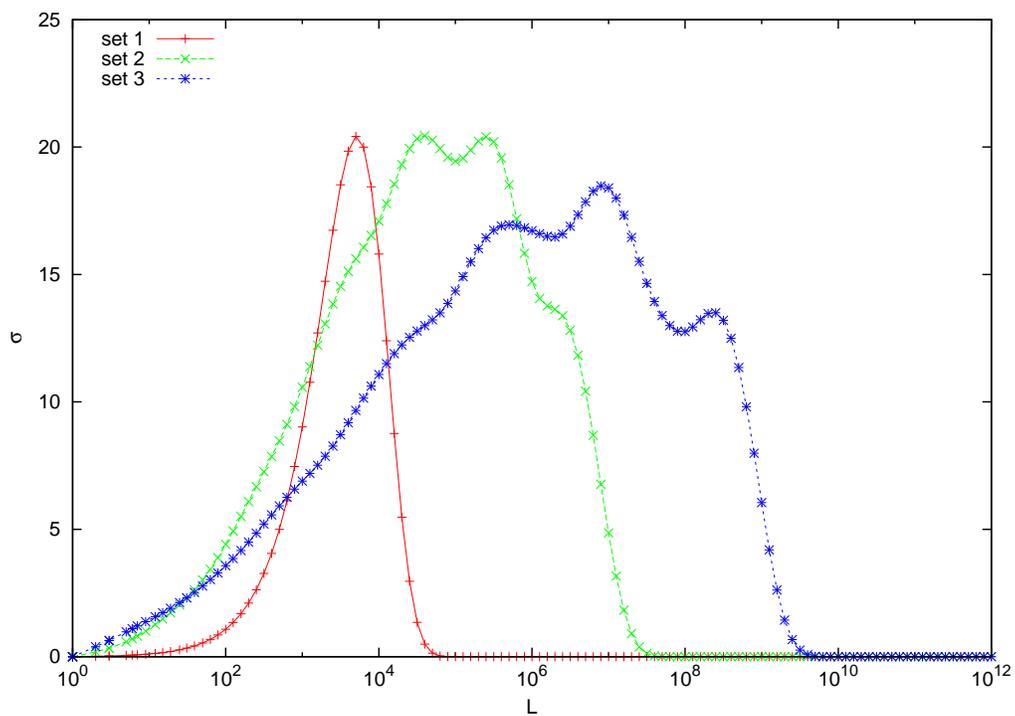}
	    }
  \caption{
    The expectation and standard deviation (the square root of the variance)
    of the number of unique sequences in the library with mutations
    in two amino acids, for three different nucleotide mixture ratios.
  \label{F:2aa}
  }
\end{figure}

\begin{figure}[!tpb]
  \centering
  \subfigure[The expectation of the number of unique sequences in the library.
  ]{
    \includegraphics[angle=270,width=\columnwidth]{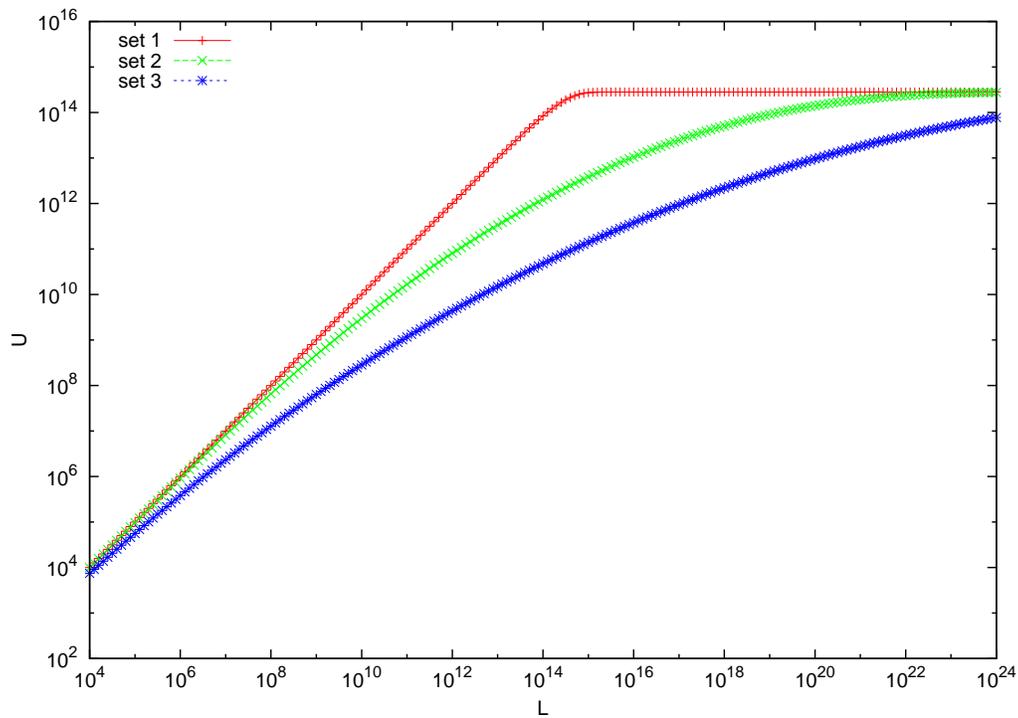}
  }
  \subfigure[The standard deviation.]
  {
    \includegraphics[angle=270,width=\columnwidth]{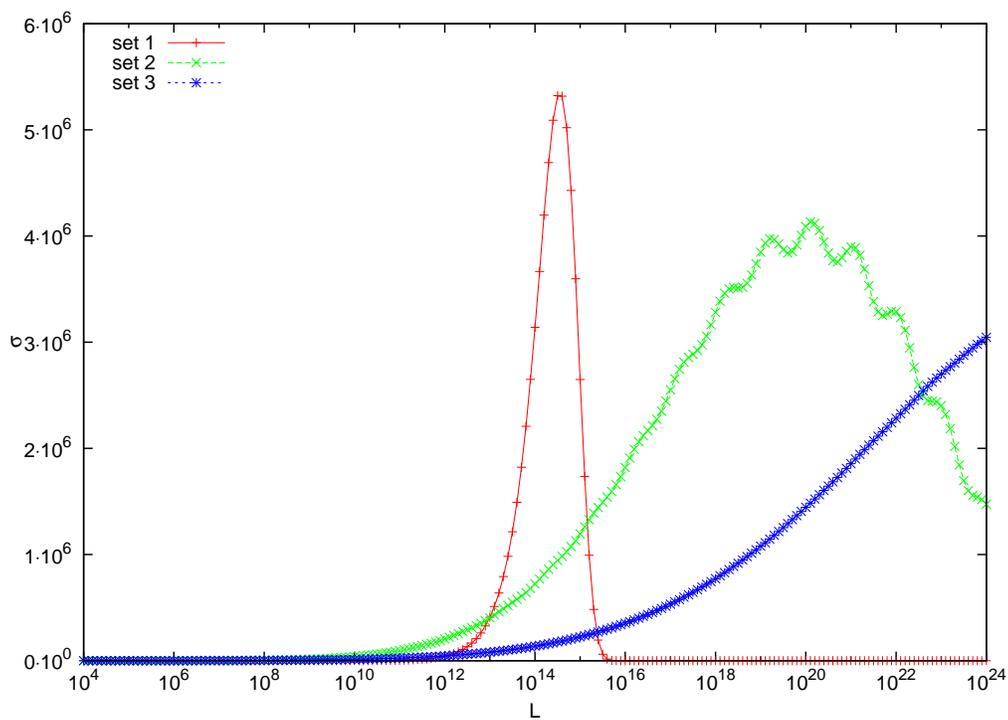}
  }
  \caption{
    The expectation and standard deviation (the square root of the variance)
    of the number of unique sequences in the library with mutations
    in $8$ amino acids, for three different nucleotide mixture ratios.
  \label{F:8aa}
  }
\end{figure}

The average and standard deviation (square root of
the variance) of the number of unique sequences in the libraries
are shown in Figures~\ref{F:2aa} and \ref{F:8aa}
as a function of $\lib$, the size of the library 
(or the number of transformants).
From these figures we can see that the nucleotide ratios of the mixtures
have a significant impact on the statistics.
As expected,
for both small and large libraries,
libraries with equimolar ratios achieve a larger number of unique 
library members $U$ with the same library size $L$ 
than libraries with unequal ratios.
On average it requires more transformants for the libraries 
with unequal ratio mixtures 
to include the same number of unique sequences.
For example, for the small library shown in Figure~\ref{F:2aa},
the library with equal ratio mixtures (Set 1) will have an average number of
unique sequences $4096$ (the theoretical limit) at $\lib \approx 10^4$,
while the library with unequal ratio mixtures Set 2 needs a larger 
library size $\lib \approx 10^7$ to get the same average number of
unique sequences,
and the library with a more skewed 
ratio mixtures Set 3 needs $\lib \approx 10^9$.
However, the ability to calculate precise library statistics allows 
one to more accurately assess the disadvantages of reduced library complexity
against the advantages of selection for particular amino acids 
(e.g. hydrophobic amino acids) by using skewed nucleotide ratios.
It should be mentioned that the nucleotide complexity 
considered here is not necessarily the same thing as amino acid 
complexity, due to the codon degeneracy.

The standard deviation also behaves differently according to the
difference in mixture ratios.
For equal ratio, the standard deviation has a sharp peak,
centered around the point where the mean is about $70\%$ of the satuation.
For unequal ratios, however,
the distributions of standard deviation is broad and multimodal,
and the peaks shift to the higher $\lib$.
For large libraries, the standard deviation is quite small when
compared with the mean. 

The number of peaks in the standard deviation depends on the
number of distinct sequence probabilities $p_i$.  
In fact, if all $n$ probabilities are distinct, there will be
$n-1$ peaks in the standard deviation.
If these peaks from the left to the right are labeled from $1$ to $n-1$,
then the peak $i$ is associated with the fluctuation 
of the number of unique sequences 
in the transition
from $\U_n = i$ to $\U_n = i+1$ as $\lib$ increases.  
When some or all sequence probabilities become degenerate, the peaks
in the standard deviation will merge together.  In the extreme case of
equal ratios, all peaks merge into one peak and the standard deviation
becomes single-modal.

\section*{Acknowledgment}
  This work was supported by
  Yale School of Medicine.
  The author would like to thank Dr. Daniel DiMaio and Sara Marlatt
  for bringing this problem to his attention.

\bibliographystyle{natbib}

\end{document}